\documentclass[11pt]{article}

\usepackage{amssymb}

\usepackage{color,cite,graphicx}
\usepackage{latexsym,amssymb,epsf} %

\usepackage{amstext}
\usepackage{amsfonts}
\usepackage{amsmath}

\newtheorem{theorem}{Theorem}[section]    
\newtheorem{fact}[theorem]{Fact}    
\newtheorem{corollary}[theorem]{Corollary}    
\newtheorem{lemma}[theorem]{Lemma}    
\newcommand{\qed}{\hfill{$\rule{6pt}{6pt}$}} 
\newenvironment{proof}{\noindent{\bf Proof:}}{\qed}
\newenvironment{proofof}[1]{\noindent{\bf Proof of #1:}}{\qed}

\newcommand{\defeq}{\stackrel{\Delta}{=}}
\newcommand{\Tr}{{\mathrm{Tr}}}

\newcommand{\complex}{{\mathbb C}}
\newcommand{\Exp}{{\mathbb{E}}}
\newcommand{\ket}[1]{| #1 \rangle}
\newcommand{\density}[1]{| #1 \rangle\!\langle #1 |}
\newcommand{\trnorm}[1]{\| #1 \|_{\mathrm t}}
\newcommand{\trace}{{\mathrm{Tr}}}

\newcommand{\ketbra}[1]{| #1 \rangle \langle #1 |}
\newcommand{\braket}[2]{\langle #1 | #2 \rangle }
\newcommand{\set}[1]{\left\{ #1 \right\} }

\newcommand{\cE}{{\mathcal{E}}}
\newcommand{\cH}{{\mathcal{H}}}
\newcommand{\cK}{{\mathcal{K}}}
\newcommand{\cM}{{\mathcal{M}}}

\newcommand{\linear}{{\mathrm{L}}}

\newcommand{\rH}{{\mathrm{H}}}
\newcommand{\rS}{{\mathrm{S}}}
\newcommand{\rI}{{\mathrm{I}}}
\newcommand{\rQ}{{\mathrm{Q}}}
\newcommand{\rB}{{\mathrm{B}}}

\newcommand{\e}{{\mathrm{e}}}

\newcommand{\acc}{{\mathrm{I}_{\mathrm{acc}}}}

\newcommand{\suppress}[1]{}

\title{\bf Accessible versus {H}olevo Information for a Binary Random
Variable}

\author{ 
Rahul Jain \\ U.\ Waterloo \thanks{School of Computer Science, and
Institute for Quantum Computing,
University of Waterloo, 200 University Ave.\ W., Waterloo, ON N2L 3G1,
Canada. Email: {\sf rjain@cs.uwaterloo.ca}. This work was
done while the author was at University of California, Berkeley,
CA~94720, USA. This work was supported by an Army Research Office
(ARO), North California, grant number DAAD 19-03-1-00082.  }
\and
Ashwin Nayak \\ 
U.\ Waterloo \& Perimeter \thanks{
Department of Combinatorics and Optimization, and Institute for
Quantum Computing, University of Waterloo, 200 University Ave.\ W.,
Waterloo, ON N2L 3G1, Canada.
E-mail: {\sf anayak@math.uwaterloo.ca}. 
A.N.\ is also Associate Member, Perimeter Institute for Theoretical
Physics, Waterloo, Canada. His research is supported in part by
NSERC, CIFAR, MITACS, CFI, OIT, and an ERA (Canada).
Research at Perimeter Institute for Theoretical Physics is supported
in part by the Government of Canada through NSERC and by the Province
of Ontario through MRI.
} 
}

\date{September 26, 2007}

\begin{document}

\maketitle

\begin{abstract}
The {\em accessible information\/}~$\acc(\cE)$ of an ensemble~$\cE$
is the maximum mutual information between a random variable encoded
into quantum states, and the probabilistic outcome of a quantum
measurement of the encoding. Accessible information is extremely
difficult to characterize analytically; even bounds on it are hard to
place. The celebrated {\em Holevo bound\/} states that accessible
information cannot exceed~$\chi(\cE)$, the quantum mutual information
between the random variable and its encoding. However, for general ensembles,
the gap between the~$\acc(\cE)$ and~$\chi(\cE)$ may be arbitrarily
large.

We consider the special case of a binary random variable, which often
serves as a stepping stone towards other results in information theory
and communication complexity.  We give explicit lower bounds
on the the accessible information~$\acc(\cE)$ of an ensemble~$\cE
\defeq \{(p, \rho_0), (1-p, \rho_1)\}$, with~$0 \leq p \leq 1$, as
functions of~$p$ and~$\chi(\cE)$.  The bounds are incomparable in the
sense that they surpass each other in different parameter regimes.

Our bounds arise by measuring the ensemble according to a complete
orthogonal measurement that preserves the fidelity of the
states~$\rho_0,\rho_1$.  As an intermediate step, therefore, we give
new relations between the two quantities~$\acc(\cE), \chi(\cE)$ and
the fidelity~$\rB(\rho_0,\rho_1)$.
\end{abstract}

\section{Introduction}
\label{sec-intro}

Let~$X$ be a classical random variable taking values in a finite
set~$\{0, \ldots, n-1 \}$ such that $\Pr(X = i) \defeq p_i$. Let~$M$
be an encoding of~$X$ into (possibly mixed) quantum states in a finite
dimensional, say $d$-dimensional, Hilbert space~$\complex^d$, such
that~$M = \rho_i$ when~$X = i$. This gives rise to an ensemble of
quantum states~$\cE \defeq \{(p_i, \rho_i) \}$. 

The mapping~$i \mapsto \rho_i$ may be viewed as a quantum
communication channel, and it is natural to ask how much information
about~$X$ can be obtained from the transmitted signal~$M$. The answer
to this question depends heavily on the way we quantify the notion of
``information''. For example, one may seek to maximize the probability
of guessing, via a measurement, the value~$i$ given an unknown
state~$\rho_i$ from the ensemble~$\cE$~\cite{Helstrom76}. This quantity
frequently arises in quantum communication, but has no simple
description in terms of the ensemble.  For a boolean random variable,
the answer is related to the trace distance of the two density
operators~\cite[pp.~106--108]{Helstrom76}.  While no analytical
expression for this probability is known in the general case, we can
still place meaningful bounds on it (see, e.g., Ref.~\cite{NayakS06}).

A different way of quantifying the information content of an ensemble
arises in Quantum Information Theory. Consider a classical random
variable~$Y^\cM$ that represents the result of a measurement of the
encoding~$M$ according to the POVM~$\cM$. The {\em accessible
  information\/}~$\acc(\cE)$ of the ensemble~$\cE$ is defined as the
maximum mutual information~$\rI(X:Y^\cM)$ obtainable via a quantum
measurement~$\cM$:
\begin{eqnarray}
\label{eqn-acc}
\acc(\cE) & \defeq & \max_{\text{POVM } \cM} \quad \rI(X:Y^\cM).
\end{eqnarray}
Accessible information is extremely difficult to characterize
analytically, even for a binary random variable (see, e.g.,
Ref.~\cite[page~1222]{FuchsG99}, where it is referred to as {\it Shannon
Distinguishability\/}).  In a celebrated result,
Holevo~\cite{Holevo73} bounded the accessible information for an
ensemble~$\cE$ by the quantum mutual information between the random
variable~$X$ and its encoding~$M$:
\begin{eqnarray}
\label{eqn-holevo}
\acc(\cE) &  \leq & \chi(\cE) 
    \quad \defeq \quad \rS(\Exp_i[\rho_i]) - \Exp_i[\rS(\rho_i)] 
    \quad = \quad \rI(X:M),
\end{eqnarray}
where~$\rS(\rho)$ denotes the von Neumann entropy of a density
matrix~$\rho$, and~$\rI(A:B) = \rS(A) + \rS(B) - \rS(AB)$ denotes the
mutual information of a bipartite quantum system~$AB$, and~$\Exp_i$
denotes the expectation (i.e., average), taken according to the
distribution~$\set{p_i}$ in the ensemble~$\cE$.  The
quantity~$\chi(\cE)$ has come to be called the {\it Holevo
  information\/} of the ensemble.

The Holevo bound is attained by ensembles of commuting states
(equivalently, for classical mixtures). In fact,
Ruskai~\cite[Section~VII.B]{Ruskai02} (see also
Ref.~\cite[Section~4.3]{Petz03}) proves that these are the only
ensembles for which~$\rI(X:Y^\cM) = \rI(X:M)$ is possible for some
fixed measurement.  A theorem due to Davies~\cite[Theorem~3]{Davies78}
establishes that there is a measurement with at most~$d^2$ outcomes
that achieves accessible information for an ensemble
of~$d$-dimensional states. (See also
Ref.~\cite[Lemma~5]{SasakiBJOH99}, and the conjecture in
Refs.~\cite{Levitin95,OsakiHB98} regarding this measurement.)
Consequently, the Holevo bound is attained only by ensembles of
commuting states.  There are ensembles for which~$\acc(\cE)$ may be
arbitrarily smaller than~$\chi(\cE)$: a uniformly random ensemble
of~$n$ states in a~$d$-dimensional space (where~$d$ is suitably
smaller than~$n$) has this property with high probability.  

\suppress{ This above fact was pointed out to us by Patrick Hayden. To
  prove this, we may use, e.g., the McDiarmid inequality to show that
  the mixture is close to completely mixed. Therefore the entropy of
  the mixture is close to~$\log d$. Using the same inequality, we can
  show that the mutual information is close to zero, with high
  probability. Another natural ensemble for which it holds are coset
  states that arise from the Graph Isomorphism problem. Here,
  individual pairs of states are far apart, however, no single
  measurement can distinguish them simultaneously.}

Lower bounds on~$\acc$ are hard to derive even for specific ensembles
(cf. Refs.~\cite{FuchsC94, JozsaRW94}). In the simplest case of a
binary random variable, Fuchs and van de Graaf~\cite{FuchsG99} relate
accessible information to the trace distance and fidelity of the
density matrices. Fuchs and Caves~\cite{FuchsC95} also consider the
case of a binary random variable, stopping short of an explicit lower
bound. It is also conjectured that the two-outcome measurement
achieving the trace-distance (or fidelity) between two pure states
occurring with equal probability also achieves accessible
information~\cite{Levitin95}. This was numerically verified, but not
formally proven, by Osaki {\em et al.\/}~\cite{OsakiHB98}. We revisit
this special case, and give a lower bound for accessible information
in terms of Holevo information. We show the following:
\begin{theorem}
\label{thm-main}
Let $\cE \defeq \{(p, \rho_0), (1-p, \rho_1)\}$ be an binary ensemble
over quantum states of any (finite) dimension. Then
\begin{eqnarray}
\label{part1}
\acc(\cE) & \geq & \rH(p)- \sqrt{4p(1-p)  - \chi(\cE)^2},
    \text{ and} \\
\label{part2}
\suppress{
\acc(\cE) & \geq & 
    -2p \log_2 \left( \sqrt{p} + \sqrt{1-p} \, f(\chi(\cE),p) \right)
    -2(1-p) \log_2 \left( \sqrt{1-p} + \sqrt{p} \, f(\chi(\cE),p) \right), 
     \\
\label{part3}
}
\acc(\cE) & \geq & 
    -\log_2 \left[ p^2 + (1-p)^2 +
    2 p(1-p) \, \left\{ 1 - \frac{\chi(\cE)^2}{4p(1-p)}
    \right\}^{1/2} \right].
\end{eqnarray}
\end{theorem}
These lower bounds arise from a specific kind of measurement of the
quantum states, namely, a measurement that preserves
fidelity~\cite{FuchsC95}. As an intermediate step, we therefore relate
both accessible information and Holevo information to fidelity in
different ways so as to give the stated bounds.

Note that the first bound above, Eq.~(\ref{part1}),
may be negative for certain parameter regimes. For
example, the bound is negative for an ensemble of
two pure states at an angle of~$\pi/4$, as the probability~$p$
approaches~$0$. 
This occurs because our measurement completely disregards the prior
probability~$p$, and because of the approximations we make in order to
get an explicit relationship with Holevo information. The second bound
is meaningful for all parameter regimes. On the other hand, for any two
orthogonal states, the first bound is greater than the second for~$p
\not\in \set{0,1,1/2}$.
\suppress{
Sharper versions of these
bounds are derived in Lemma~\ref{lem:bleqacc} in terms of the
fidelity. We compare them further at that point.
}

The motivation for this work is to investigate the extent to
which the Holevo bound~$\acc(\cE) \leq \chi(\cE)$ is tight for binary
ensembles. Note also that~$\chi(\cE) = \rI(X:M) \leq \rH(p)$ by the
Lanford-Robinson inequality~\cite[Eq.~(2.3), page~238]{Wehrl78}
(also~\cite[Theorem~11.10, page~518]{NielsenC00}). Our main theorem
may thus be seen as a sharpening of these bounds. None of the
previously known bounds provide a lower bound on accessible
information as an explicit function of Holevo information.

Relations between different measures of information in the binary case
often form a stepping stone in results in information theory. For
example, as explained in Ref.~\cite{FuchsC94}, the Holevo bound may be
derived by studing the case of the binary ensemble. In complexity
theory, these relations form the basis of strong, sometimes optimal,
lower bounds for the quantum communication complexity of computing
functions, as in Refs.~\cite{KlauckNTZ07, JainRS03b}. They have also
found application in the study of quantum interactive proof
systems~\cite{KitaevW00} and quantum coin
flipping~\cite{Ambainis04,SpekkensR02a}.  We expect that our
inequalities provide a more operationally useful view of accessible
information, and find similar application.

\section{Comparison with previous bounds}
\label{sec-comparison}

A number of lower bounds on accessible information~$\acc(\cE)$ are
already known. Jozsa, Robb, and Wootters~\cite{JozsaRW94} bounded it
for arbitrary (possibly non-binary) ensembles by the expected mutual
information resulting from a uniformly random complete orthogonal
measurement. The latter quantity depends solely on the average density
matrix~$\rho = \Exp_i \rho_i$ corresponding to the ensemble, and was
defined as the {\em subentropy\/}~$\rQ(\rho)$ of the density
matrix~$\rho$. They also showed that for every density matrix~$\rho$
(of arbitrarily large finite dimension), subentropy expressed in bits
is bounded as~$\rQ(\rho) \leq (1 - \gamma) \log_2 \e \leq 0.60995$,
where~$\gamma$ is the Euler constant. Thus, the best possible lower
bound resulting from the subentropy bound is~$0.60995$. Our bound,
Theorem~\ref{thm-main} (Eq.~(\ref{part1})), is more sensitive to the
ensemble. For example, it gives the sharper lower bound of~$1 -
\sqrt{2\epsilon}$ for an equally weighted binary ensemble~$\cE$ for
which~$\chi(\cE) = 1 - \epsilon$, for any~$\epsilon \in
[0,1]$. Holevo~$\chi$ information arbitrarily close to~$1$ is
achieved, among others, by a uniform distribution over two pure states
whose inner product is arbitrarily close to~$0$. Thus our bound better
reflects the ensemble-dependence of accessible information.

Fuchs and Caves~\cite{FuchsC94} considered a measurement of a binary
ensemble~$\cE$ that arises in one proof of the Holevo bound.
This is a measurement~$\cM$ that minimizes the second
derivative of the mutual information~$I(X:Y^\cM)$ (as defined in 
Section~\ref{sec-intro})
with respect to the variable~$p$. The measurement may be computed
numerically given any ensemble, and leads to a lower bound on accessible
information for binary ensembles.  In a few special cases with single qubit
states, such as when the states are pure, or commute, the 
Fuchs-Caves measurement has a
familiar description. In all these special cases, the measurement
is in a basis that diagonalizes~$p\, \rho_0 - (1-p) \, \rho_1$;
precisely the measurement that maximizes the probability of
guessing~$i$ correctly given a state~$\rho_i$ picked from the
ensemble~$\cE$. This measurement also preserves fidelity between the
states~$\rho_0, \rho_1$ when~$p = 1/2$, and the lower bound with which
we start matches this (see the proof of Lemma~\ref{lem:bleqacc}).  Our
final lower bounds are however weaker, since we make further
approximations in order to relate accessible information to the
Holevo~$\chi$ quantity. For~$p \not= 1/2$, even our initial lower
bound may be worse, since fidelity preserving measurements are
independent of the value of~$p$.

Hall~\cite[Eq.~24, page~104]{Hall97} considered the lower bound on
accessible information for arbitrary (possibly non-binary) ensembles
that arises from the ``pretty good
measurement''. A special case of this measurement, when the ensemble
consists of pure states, was defined by Hughston, Jozsa, and
Wootters~\cite{HughstonJW93}, later named so
and studied by Hausladen and Wootters~\cite{HausladenW94}. The
measurement may be viewed as a dual to ensembles, and is a natural
candidate for state discrimination. The bound that arises from the
pretty good measurement may also be computed numerically for any given
ensemble, as with the previous two bounds.
In all the cases of an ensemble with two-states that we studied
numerically, the Hall bound follows the Fuchs and Caves bound very
closely. We are not aware of any formal connection between the two,
though. Our numerical findings differ from the ones originally
presented by Hall, and have been verified by him in a private
communication. (Figure~1 in~\cite[p.~104]{Hall97} is therefore
in error.)

Typically for applications in complexity theory and cryptography such as
those mentioned in Section~\ref{sec-intro}, we seek simple functional 
relationships between different measures of information.
All the three bounds discussed here can be evaluated numerically on a
computer but seem not to lead to an explicit functional connection
with Holevo-$\chi$ information. Our work attempts to fill this gap.

\begin{figure}[t]
\begin{center}
\includegraphics[height=10cm,angle=-90]{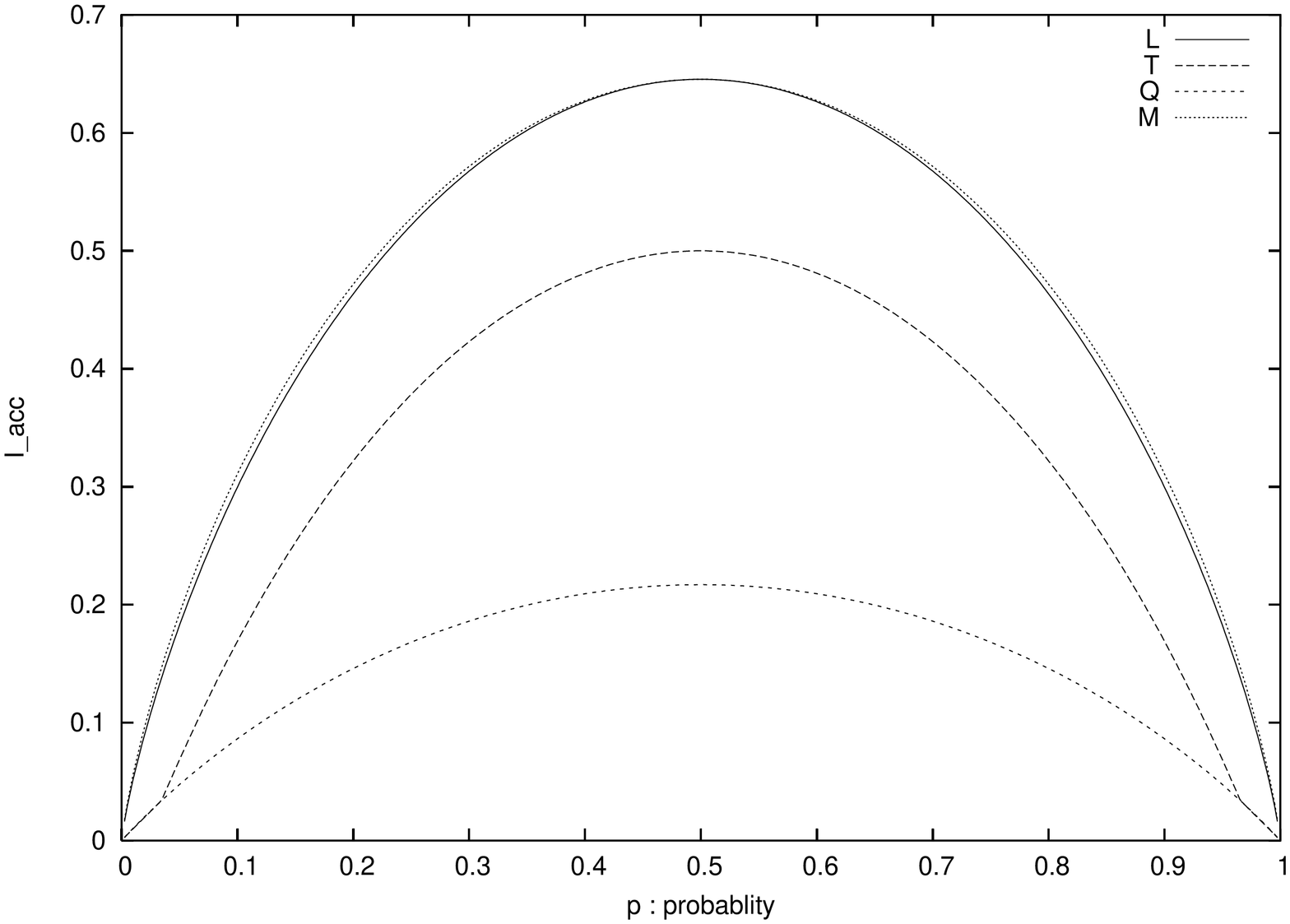}
\caption{Bounds for an ensemble~$\cE = \set{ (p,\rho_0),
(1-p,\rho_1)}$ of two pure states with inner-product~$1/2$. The bounds
on~$\acc(\cE)$ are plotted as~$p$ varies in~$[0,1]$.}
\label{fig-1}
\end{center}
\end{figure}

\begin{figure}
\begin{center}
\includegraphics[height=10cm,angle=-90]{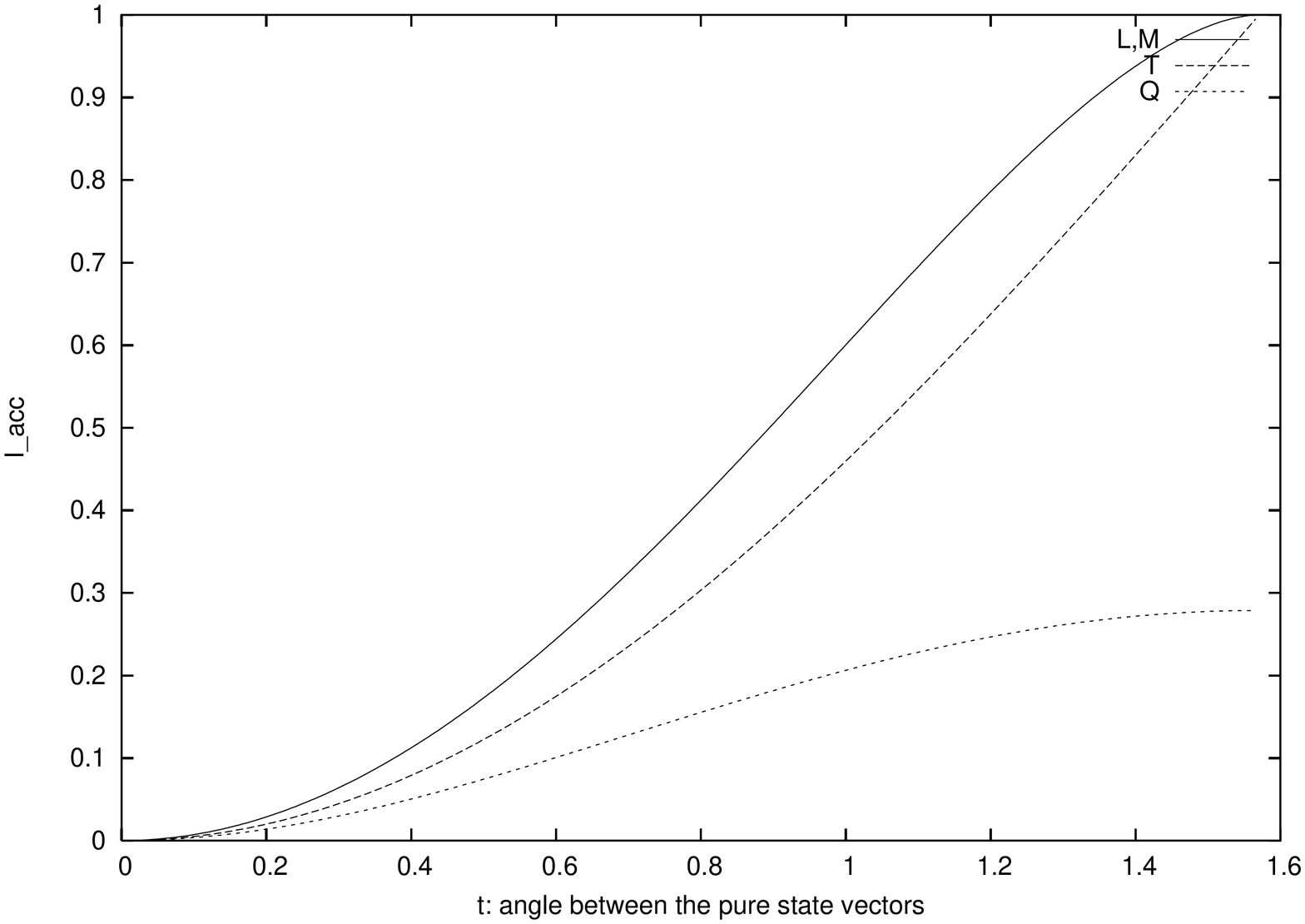}
\caption{Bounds for a uniform ensemble~$\cE = \set{ (1/2,\rho_0),
(1/2,\rho_1)}$ of two pure states at an angle~t with each other, as~t
varies in~$[0,\pi/2]$.}
\label{fig-2}
\end{center}
\end{figure}

\begin{figure}
\begin{center}
\includegraphics[height=10cm,angle=-90]{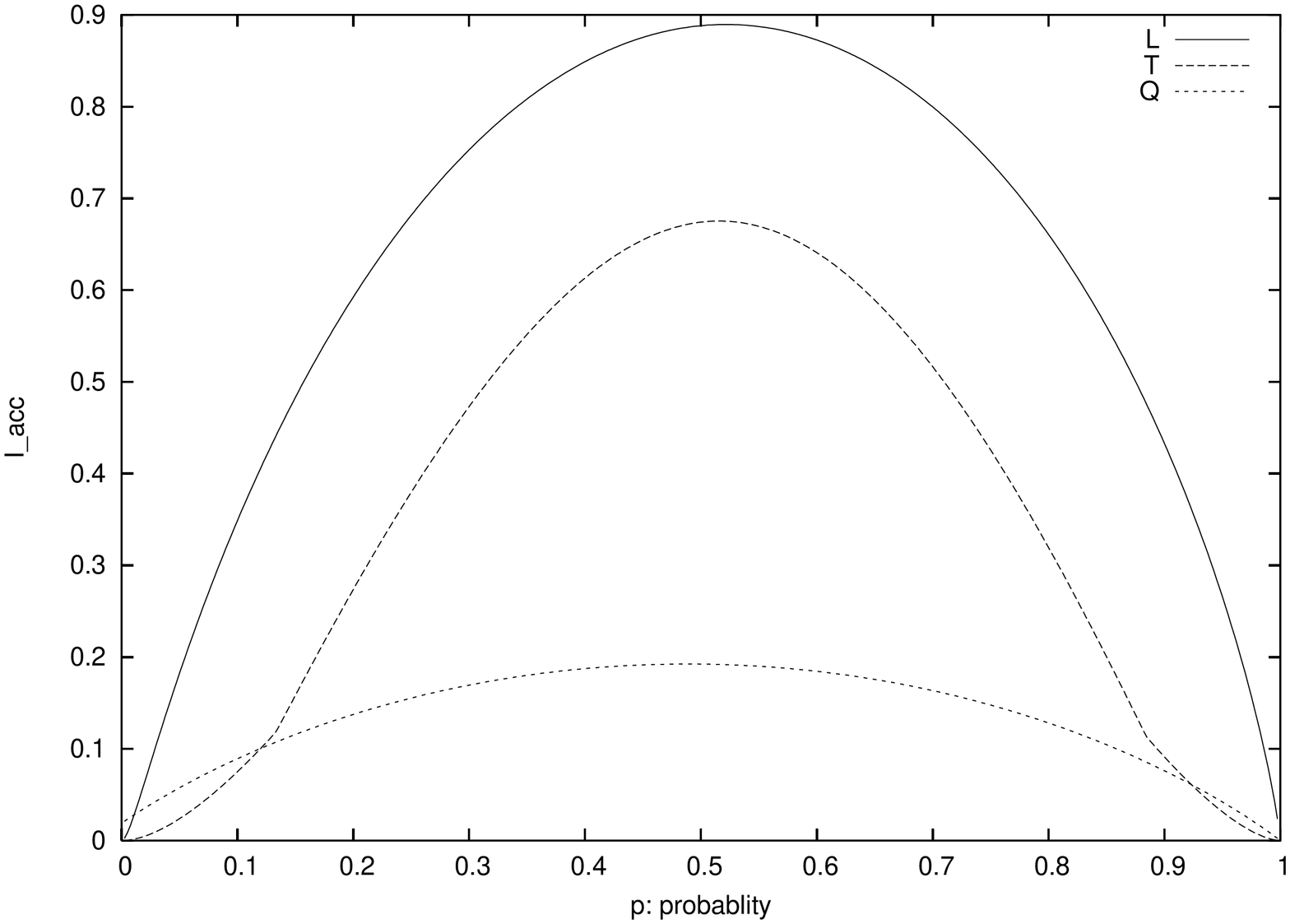}
\caption{Bounds for an ensemble~$\cE = \set{ (p,\rho_0),
(1-p,\rho_1)}$ of a mixed qutrit state~$\rho_0= 0.01 \density{0} +
0.01 \density{1} + 0.98 \density{2}$ and a pure qutrit state~$\rho_1 =
\density{v}$, where~$v = \sqrt{0.02}\ket{0} + \sqrt{0.96}\ket{1} +
\sqrt{0.02}\ket{2}$. The bound~M is not plotted here.  The
bounds on~$\acc(\cE)$ are plotted as~$p$ varies in~$[0,1]$.}
\label{fig-3}
\end{center}
\end{figure}

In Figures~\ref{fig-1}--\ref{fig-3}, we present a numerical comparison
of the abovementioned bounds with the maximum of our lower bounds,
which is denoted by~T.  The Jozsa {\it et al.\/} bound is marked as~Q,
the Fuch-Caves bound is marked as~M, and the Hall bound is marked
as~L. In all the cases we plotted, our bound~T is below~M
and~L. Except when the probability~$p$ is close to~$0$ or~$1$ in
Figure~\ref{fig-3}, T is well above~Q. For pure state ensembles
with~$p = 1/2$, the lower bound
with which we start in Lemma~\ref{lem:bleqacc} equals~L
and~M. However, our final bound~T is an approximation to this bound,
and is lower (see Figure~\ref{fig-2}).

\section{Lower bounds on accessible information}

\subsection{Preliminaries}

We quickly summarize some information theoretic concepts and notation
we use in this article. For a more comprehensive treatment, we refer
the reader to a text such as~\cite{NielsenC00}.

All logarithms in this article are taken to the base~$2$. For $0 \leq
p \leq 1$, $\rH(p) \defeq -p \log p - (1-p) \log (1-p)$ denotes the
binary entropy function.  The following fact (see,
e.g.,~\cite[page~1224, Fig.~1]{FuchsG99}) bounds the binary entropy
function close to~$1/2$:
\begin{fact}
\label{fact:upph}
For $\delta \in [-\frac{1}{2}, \frac{1}{2}], \quad \rH(\frac{1}{2} + \delta)
\leq \sqrt{1 - (2\delta)^2}$. 
\end{fact}

Let $\cH, \cK$ be Hilbert spaces. For a quantum state~$\rho \in
\linear(\cH)$, we call a pure state~$\ket{\phi} \in \cH \otimes \cK $
a {\em purification\/} of $\rho$ if $\Tr_{\cK}\, \ketbra{\phi} = \rho
$. The {\em fidelity} between (mixed) quantum states $\rho, \sigma $
is defined as $\rB(\rho, \sigma) \defeq \sup\; |\braket{\phi}{\psi}|
$, where the optimization is over states $\ket{\phi}, \ket{\psi}$
which are purifications of $\rho$ and $\ket{\psi}$, respectively.  The
{\em trace norm\/} of an operator $A \in \linear(\cH)$ is defined as
$\trnorm{A} \defeq \Tr\, \sqrt{A^{\dagger}A}$.

The fidelity of two mixed states may be expressed in terms of the
trace norm.
\begin{theorem}[Uhlmann~\cite{Uhlmann76,Jozsa94}]
\label{thm:fidelity-trace}
For any quantum states~$\rho$ and~$\sigma$, 
$\rB(\rho,\sigma) = \trnorm{\sqrt{\rho}\sqrt{\sigma}}$.
\end{theorem}

The {\em von-Neumann\/} entropy of a quantum state $\rho$ is defined
as $\rS(\rho) \defeq - \Tr\; \rho \log \rho $. The {\em
Kullback-Leibler divergence\/} or {\em relative entropy\/} between two
quantum states $\rho, \sigma$ is defined as $\rS(\rho \| \sigma)
\defeq \Tr\; \rho \, (\log \rho - \log \sigma)$. The following fact
relates fidelity and relative entropy:
\begin{lemma}[Dacunha-Castelle~\cite{Dacunha-Castelle78}]
\label{lem:bleqs} 
For any quantum states~$\rho,\sigma$, we have~$\rS(\rho \| \sigma)
\geq -2 \log \rB(\rho, \sigma)$.
\end{lemma}

The following relation follows from the definitions:
\begin{lemma}
\label{lem:relinf} 
Let $X$ be a random variable over a finite sample space and let $p_x
\defeq \Pr(X=x)$. Let $M$ be a quantum encoding of $X$ such that $M =
\rho_x$ when $X = x$. Let $\rho \defeq \Exp_x[\rho_x] = \sum_x p_x
\rho_x$.  Then $\rI(X:M) \; = \; \Exp_x[\rS(\rho_x \| \rho)]$.
\end{lemma}

\subsection{The lower bounds}

We begin with a property of fidelity that we later use.
\begin{lemma}
\label{lem:ub2}
Let~$\cE \defeq \{(p, \rho_0), (1-p, \rho_1)\}$ be an ensemble of
commuting quantum states of arbitrary (finite) dimension, and
let~$\rho = p\, \rho_0 + (1-p)\, \rho_1$. Then
\begin{eqnarray*}
\suppress{
\label{eqn-ub1}
\rB(\rho_0,\rho) & \leq & \sqrt{p} + \sqrt{1-p} \; \rB(\rho_0,\rho_1),
\quad \text{and} \\
\label{eqn-ub2}}
p \, \rB(\rho_0,\rho) + (1-p) \, \rB(\rho_1,\rho) & \leq & \left[ p^2 +
(1-p)^2 + 2p(1-p) \, \rB(\rho_0,\rho_1) \right]^{1/2}.
\end{eqnarray*}
\end{lemma}
\begin{proof}
For any two commuting positive semi-definite matrices~$A,B$, we have
\begin{eqnarray*}
\suppress{
\sqrt{A+B} & \leq & \sqrt{A} + \sqrt{B}.
}
\trnorm{AB} & = & \trace \; AB.
\end{eqnarray*}
Since~$\rho_0,\rho_1,\rho$ all commute, we have
\suppress{
\begin{eqnarray*}
\rB(\rho_0,\rho) & = & \trnorm{ \sqrt{\rho_0} \sqrt{\rho}} \\
    & = & \trace \;  \left[ \sqrt{\rho_0} \sqrt{\rho} \right] \\
    & \leq & \trace \; \left[ \sqrt{\rho_0} \sqrt{ p \, \rho_0} +
             \sqrt{\rho_0} \sqrt{ (1-p)\, \rho_1} \right] \\
    & = & \sqrt{p} + \sqrt{1-p} \, \trnorm{\sqrt{\rho_0}\sqrt{\rho_1}},
\end{eqnarray*}
which is the first inequality.

Similarly, we have
}
\begin{align*}
\lefteqn{p \, \rB(\rho_0,\rho) + (1-p) \, \rB(\rho_1,\rho)} \\
    & = \quad p \, \trnorm{ \sqrt{\rho_0} \sqrt{\rho} }
          + (1-p) \, \trnorm{ \sqrt{\rho_1} \sqrt{\rho} }
        & \text{By Theorem~\ref{thm:fidelity-trace}} \\
    & = \quad p \, \trace \left[ \sqrt{\rho_0} \sqrt{\rho} \right]
          + (1-p) \, \trace \left[ \sqrt{\rho_1} \sqrt{\rho} \right] 
        &    \\
    & = \quad \trace \left[ \left(  p \sqrt{\rho_0} 
          + (1-p)\sqrt{\rho_1} \right) \cdot \sqrt{\rho} \right] 
        & \\
    & \leq \quad \left[ \trace \; \left(  p \sqrt{\rho_0} 
          + (1-p)\sqrt{\rho_1} \right)^2 \right]^{1/2}
          \cdot \left[ \trace \; \rho \right]^{1/2}, 
        & \text{By Cauchy-Schwartz} \\
    & = \quad \left[ p^2 + (1-p)^2 + 2p(1-p) \; \trace
    \left(\sqrt{\rho_0}\sqrt{\rho_1}\right) \right]^{1/2}, &
\end{align*}
which is the inequality we seek.
\end{proof}

We now bound the accessible information from below by the mutual
information achieved by a measurement that preserves fidelity.
\begin{lemma}
\label{lem:bleqacc}
Let $\cE \defeq \{(p, \rho_0), (1-p, \rho_1)\}$ be an ensemble of
quantum states of arbitrary (finite) dimension. Then
\begin{eqnarray}
\label{eqn-lb1}
\acc(\cE) & \geq & \rH(p) - 2\sqrt{p(1-p)}\; \rB(\rho_0,\rho_1),
    \text{ and} \\
\label{eqn-lb2}
\suppress{
\acc(\cE) & \geq & 
    -2p \log_2 \left( \sqrt{p} + \sqrt{1-p} \,\rB(\rho_0,\rho_1) \right)
    -2(1-p) \log_2 \left( \sqrt{1-p} + \sqrt{p} \,\rB(\rho_0,\rho_1) \right), 
                   \\
\label{eqn-lb3}
}
\acc(\cE) & \geq & 
    -\log_2 \left[ p^2 + (1-p)^2 + 2p(1-p) \, \rB(\rho_0,\rho_1) 
    \right].
\end{eqnarray}
\end{lemma}
\begin{proof}
Let $\rho_0', \rho_1'$ be the classical distributions resulting from a
measurement that achieves the fidelity between $\rho_0$ and $\rho_1$
(see Ref.~\cite{FuchsC95}), so that~$\rB(\rho_0',\rho_1') =
\rB(\rho_0,\rho_1)$. Let~$\rho' = p \rho_0' + (1 -p) \rho_1'
$.  Let $M'$ be the encoding of $X$ such that $M' = \rho_0'$ when
$X=0$ and $M' = \rho_1'$ when $X=1$.  Let
\begin{eqnarray*}
q_0(m) & \defeq & \Pr (M'=m \,|\, X=0) \\
q(m)   & \defeq & \Pr(M'=m), \text{ and} \\
r_0(m) & \defeq & \Pr(X=0 \,|\, M'=m).
\end{eqnarray*}
We similarly define $q_1(m)$. 
Note that
\begin{equation}
\label{eqn-cond-prob}
q(m) \, r_0(m)   = p \, q_0(m) \quad \text{and} \quad 
q(m) \, (1-r_0(m)) = (1-p)\, q_1(m).
\end{equation}

We now have
\begin{align*}
\acc(\cE) \quad
    & \geq \quad  \rI(X:M') & \\ 
    & = \quad    \rH(p) - \sum_m q(m) \, \rH(r_0(m)) & \\ 
    & \geq \quad  \rH(p) - \sum_m q(m) \cdot 2 \sqrt {r_0(m) \, (1 - r_0(m)) }
              & \text{From Fact~\ref{fact:upph}} \\
    & = \quad    \rH(p) - \sum_m 2 \sqrt{p(1-p) \, q_0(m) \, q_1(m)}
              &  \text{From Eq.~(\ref{eqn-cond-prob})} \\
    & = \quad     \rH(p) - 2\sqrt{p(1-p)} \; \rB(\rho_0', \rho_1'). &
\end{align*}
This proves the bound in Eq.~(\ref{eqn-lb1}).

\suppress{
For the second part, Eq.~(\ref{eqn-lb2}), we have
\begin{align*}
\acc(\cE) \quad
    & \geq \quad \rI(X:M') & \\
    & = \quad  p \, \rS(\rho_0' \| \rho') + (1-p) \, \rS(\rho_1' \| \rho') 
           & \text{From Lemma~\ref{lem:relinf}} \\
    & \geq \quad  -2p \, \log \rB(\rho_0', \rho') 
           - 2(1-p) \, \log \rB(\rho_1', \rho') & \text{From
           Lemma~\ref{lem:bleqs}} \\
    & \geq \quad -2p \, \log\left( \sqrt{p} + \sqrt{1-p} \; 
           \rB(\rho_0', \rho_1') \right) & \\
    & \quad\quad 
           - 2(1-p) \, \log \left( \sqrt{1-p} + \sqrt{p} \;
           \rB(\rho_0', \rho_1') \right) & \text{From
           Lemma~\ref{lem:ub2}}.
\end{align*}
Since~$\rB(\rho_0', \rho_1') = \rB(\rho_0, \rho_1)$, the bound
follows.

For the third part, Eq.~(\ref{eqn-lb3}), we begin as above, but invoke
a different set of approximations.
}

For the second part, Eq.~(\ref{eqn-lb2}), we have
\begin{align*}
\acc(\cE) \quad
    & \geq \quad \rI(X:M') & \\
    & = \quad  p \, \rS(\rho_0' \| \rho') + (1-p) \, \rS(\rho_1' \| \rho') 
           & \text{From Lemma~\ref{lem:relinf}} \\
    & \geq \quad  -2p \, \log \rB(\rho_0', \rho') 
           - 2(1-p) \, \log \rB(\rho_1', \rho') & \text{From
           Lemma~\ref{lem:bleqs}} \\
    & \geq \quad -2 \log \left( p \, \rB(\rho_0', \rho') + (1-p) \,
           \rB(\rho_1', \rho') \right) & \text{By convexity} \\
    & \geq \quad -\log \left( p^2 + (1-p)^2 + 
           2p(1-p) \, \rB(\rho_0', \rho_1') \right) & 
           \text{By Lemma~\ref{lem:ub2}}
\end{align*}
Since~$\rB(\rho_0', \rho_1') = \rB(\rho_0, \rho_1)$, the bound
follows.
\end{proof}

The bounds in the above lemma are incomparable, in that they surpass
each other in different parameter regimes as described in
Section~\ref{sec-intro}.

\suppress{
 As mentioned earlier, the
bounds in Eq.~(\ref{eqn-lb1}) and~(\ref{eqn-lb2}) are negative when
the states have fidelity approaching~$1$, except for a few values
of~$p$. The bound in Eq.~(\ref{eqn-lb3}) is always non-negative. When
the states are orthogonal, or nearly so, the first two bounds surpass
the third. Eq.~(\ref{eqn-lb1}) always gives a better bound than the second.
}

Next we observe that for pure quantum states, there is a direct
relationship between fidelity and the Holevo-$\chi$ quantity.
\begin{lemma}
\label{thm-pure-chi}
Let $\cE \defeq \{(p, \rho_0), (1-p, \rho_1)\}$ be
an ensemble such that $\rho_0, \rho_1$ are pure states. Then,
\[
\chi(\cE) \quad \leq \quad 2 \sqrt {p(1-p) (1- \rB(\rho_0, \rho_1)^2)}.
\]
\end{lemma}
\begin{proof}
  Let $\theta \in [0, \pi/2]$ be the angle between the pure states
  $\rho_0$ and $\rho_1$ so that $\rB(\rho_0, \rho_1) = \cos \theta $.
  Let $ \rho = p \rho_0 + (1 -p) \rho_1 $. By a direct calculation we
  see that the eigenvalues of $\rho$ are
\[
\frac{1 \pm \sqrt{1 - 4p(1-p)\sin^2 \theta}}{2}.
\]
Therefore from Fact~\ref{fact:upph} we have,
\begin{eqnarray*}
\chi(\cE) & = & \rS(\rho) \quad = \quad \rH\left(\frac{1 + \sqrt{1 
                                - 4p(1-p)\sin^2 \theta}}{2} \right)  \\ 
         & \leq &  (2\sin \theta)  \sqrt {p(1-p)} 
         \quad = \quad 2 \sqrt {p(1-p) (1- \rB(\rho_0, \rho_1)^2)}.
\end{eqnarray*}
\end{proof}

As a corollary of the above lemma, we get:
\begin{corollary}
\label{cor:ileqb}
Let $\cE \defeq \{(p, \rho_0), (1-p, \rho_1)\}$ be
an ensemble where $\rho_0, \rho_1$ may be mixed states. Then, 
\[
\chi(\cE) \quad \leq \quad 2 \sqrt{p(1-p) (1- \rB(\rho_0, \rho_1)^2)}.
\]
\end{corollary}
\begin{proof}
As before, let $\ket{\phi_0}, \ket{\phi_1}$ be purifications of
$\rho_0, \rho_1$ which achieve fidelity between the two states.  Let
us consider the encoding $M'$ of $X$ such that $M' = \ket{\phi_0}$
when $X=0$ and $M' = \ket{\phi_1}$ when $X=1$. From the strong
sub-additivity property of von Neumann entropy it follows that
$\rI(X:M) \leq \rI(X:M')$. Using Lemma~\ref{thm-pure-chi} we have
\begin{eqnarray*}
\chi(\cE) & = & \rI(X:M) \quad \leq \quad \rI(X:M') \\
     & \leq & 2 \sqrt {p(1-p) (1- \rB(\ketbra{\phi_0}, \ketbra{\phi_1})^2)} \\ 
     & = & 2 \sqrt {p(1-p) (1- \rB(\rho_0, \rho_1)^2)},
\end{eqnarray*}
as required.
\end{proof}

Finally we get our main result, Theorem~\ref{thm-main}.

\begin{proofof}{Theorem~\ref{thm-main}}
From Corollary~\ref{cor:ileqb} we have
\begin{eqnarray*}
\rB(\rho_0, \rho_1)
    & \leq & \left[ 1 - \frac{\chi(\cE)^2}{4p(1-p)} \right]^{1/2}.
\end{eqnarray*}
The inequalities follow by combining the above with those in
Lemma~\ref{lem:bleqacc}.
\end{proofof}

\section{Concluding remarks}

In Theorem~\ref{thm-main} we bounded the accessible information of an
arbitrary ensemble corresponding to a binary random variable from
below, by relating it to the Holevo~$\chi$ quantity. By a theorem of
Ruskai~\cite[Section~VII.B]{Ruskai02}, whenever the states in the
ensemble are not orthogonal (or equal), no measurement achieves Holevo
information. This also rules out the possibility of the two quantities
being equal in the limit of more and more refined measurements, since
the number of outcomes in the optimal measurement on a finite
dimensional space may be bounded by the Davies
Theorem~\cite[Theorem~3]{Davies78}. This implies that the Holevo bound
is strict for ensembles of non-orthogonal states. The significance of
our lower bounds is that they quantify the extent to which accessible
information may be smaller than Holevo information. 

In Section~\ref{sec-comparison}, we pointed out some advantages of our
bounds vis-a-vis previously known lower bounds.  Along with the
intermediate bounds we obtain in Lemma~\ref{lem:bleqacc} and
Corollary~\ref{cor:ileqb}, Theorem~\ref{thm-main} also adds to the
suite of relations between different distinguishability measures such
as those identified by Fuchs and van de Graaf~\cite{FuchsG99}. These
relations have found applications in a variety of areas---information
theory, cryptography, and communication complexity. We anticipate that
our bounds find similar application.  Finally, we believe that our
bounds may be further tightened. Due to the basic nature of the
question, the existence of a tighter bound would be of interest
regardless of potential applications.

\suppress{
A refinement of the Davies Theorem due to Sasaki {\em et
al.\/}~\cite[Lemma~5]{SasakiBJOH99} says that a measurement consisting
of~$d(d+1)/2$ rank~$1$ POVM elements achieves acessible information
for~$d$-dimensional states with real amplitudes.
Thus, for an ensemble with two pure states, a POVM with three 
elements achieves accessible information, and it is conjectured that
two outcomes suffice for uniform ensembles~\cite{Levitin95,OsakiHB98}.
If this conjecture is formally
verified, we would get a closed form expression for the accessible
information (in this special case) as a function of the fidelity of
the two states, potentially improving our results for general binary
ensembles. We leave the possibility of a tighter bound open.
}

\subsection*{Acknowledgements}

We thank Andris Ambainis, Debbie Leung, Jaikumar Radhakrishnan, Mary
Beth Ruskai and Pranab Sen for enlightening discussions, and Mary Beth
Ruskai also for pointers to relevant literature. 
We are grateful to Michael Hall for his comments on an earlier version
of this article, and for pointing out several errors.
We also thank the
anonymous referees for their suggestions for improvement.

\bibliographystyle{plain}
\bibliography{acc-info}

\end{document}